\documentclass[aps,prl,twocolumn,superscriptaddress,longbibliography]{revtex4-1}

\usepackage{subcaption}
\usepackage{amssymb}
\usepackage{amsthm}
\usepackage{amstext}                                                             
\usepackage{mathtools}
\usepackage{algorithm} 
\usepackage{algpseudocode}
\usepackage{float}
\usepackage{multirow}
\usepackage{xcolor}
\usepackage{breqn}
\usepackage[utf8]{inputenc}

\usepackage{pgfplots} 
\usepgfplotslibrary{groupplots}
\usepgfplotslibrary{external}
\DeclareUnicodeCharacter{2212}{-} 

\makeatletter
\let\cat@comma@active\@empty
\makeatother

\graphicspath{figures}

\begin{document}

\title{A Numerical Proof of Shell Model Turbulence Closure}

\author{Giulio Ortali}
\affiliation{Department of Applied Physics, Eindhoven University of Technology, Eindhoven, The Netherlands}
\affiliation{SISSA (International School for Advanced Studies), Trieste, Italy}
\author{Alessandro Corbetta}
\affiliation{Department of Applied Physics, Eindhoven University of Technology, Eindhoven, The Netherlands}
\author{Gianluigi Rozza}
\affiliation{SISSA (International School for Advanced Studies), Trieste, Italy}
\author{Federico Toschi}
\affiliation{Department of Applied Physics, Eindhoven University of Technology, Eindhoven, The Netherlands}


\begin{abstract}
\label{sec:abstract}
The development of turbulence closure models, parametrizing the influence of small non-resolved scales on the dynamics of large resolved ones, is an outstanding theoretical challenge with vast applicative relevance. We present a closure, based on deep recurrent neural networks, that quantitatively reproduces, within statistical errors, Eulerian and Lagrangian structure functions and the intermittent statistics of the energy cascade, including those of subgrid fluxes. To achieve high-order statistical accuracy, and thus a stringent statistical test, we employ shell models of turbulence. Our results encourage the development of similar approaches for 3D Navier-Stokes turbulence.
\end{abstract}

\maketitle

Turbulence is the chaotic and ubiquitous dynamics of fluids, almost
unavoidable for high velocity flows. Key to a vast number of
environmental and industrial flows~\cite{pope_2000}, 3D turbulence is
characterized by a nonlinear forward energy cascade from large scales,
where energy is injected, to smaller scales, where it is dissipated
via viscous friction~\cite{rev_cascade}.

The number of dynamically active degrees of freedom (\textit{DOFs}) in a turbulent flow is estimated to grow as $Re^{9/4}$~\cite{frisch}, where $Re$ is the Reynolds number, quantifying the turbulent intensity of the flow. Typical natural and industrial flows are characterized by a value of the Reynolds number so high to make it attractive, or even mandatory, the development of so called \textit{Large Eddy Simulation} (LES) capable of parameterizing, via \textit{subgrid closure models}, the effect of the smallest scales on the larger ones. 
However, the complex space-time correlations  of turbulence, characterized by
 intermittent dynamics and anomalous scaling laws~\cite{frisch},
challenge physical understanding and hinder the development of
accurate and reliable subgrid closure models. 

Given the practical importance of turbulence modelling, Large Eddy
Simulations have flourished ~\cite{les}, demonstrating important potential in  partially reproducing,  in specific geometries, some (statistical) features of the fully resolved Navier-Stokes turbulence. 

Due to the fact that turbulence is a (deterministic) chaotic system,
the best that any turbulence model could ever achieve is being able to
quantitatively reproduce all statistical properties. Additionally, due
to intermittency, these statistical properties require in principle the evaluation
of arbitrarily high-order space-time statistical
moments.
Therefore the ``perfect'' Large Eddy Simulations should be capable of
reproducing all statistical moments of the resolved scales, having
these indistinguishable from those computed from a filtered direct
numerical simulation of the Navier-Stokes equation. Such a demanding
requirement has never been accomplished before, making it unclear if this goal could be achieved at all. In this work we show that, within the highest statistical accuracy we could achieve, this task is indeed feasible.

Due to the massive amount of data needed to reach converged statistics of high order statistical moments, we operate in the simplified setting of a reduced dynamical system: the shell model of turbulence~\cite{shellrev}.  Shell models model the dynamics of homogeneous isotropic turbulence in Fourier space via a (small) number of complex-valued scalars $u_n$, $n=0,1,\dots, \infty$, whose magnitude represents the energy of fluctuations at representative logarithmically equispaced spatial scales with wavelength $k_n = k_0 \lambda^n$ (usually $\lambda=2$).
Shell models reproduce, using a small number of DOFs, the main features of the turbulent energy cascade, including intermittency and anomalous scaling exponents, which are practically indistinguishable from those of real 3D turbulence~\cite{sabra}.

The challenge of subgrid closure in the context of the shell model consists in resolving only a subset of the shell variables, $\{u_n, n=0,\dots, N_{cut}\}$, above an arbitrary \textit{subgrid cutoff scale} $N_{cut} \ll N_{\eta}$, where $N_{\eta}$ corresponds to the Kolmogorov scale (i.e.\ $k_{N_{\eta}}$ is the wavelenght of the Kolmogorov scale).  This entails modelling the effects of the small scales $\{u_n, n= N_{cut}+1,\dots, N_{\eta}\}$ on the large scales.

\begin{figure}[h]
   \centering \resizebox{.5\textwidth}{!}{\includegraphics{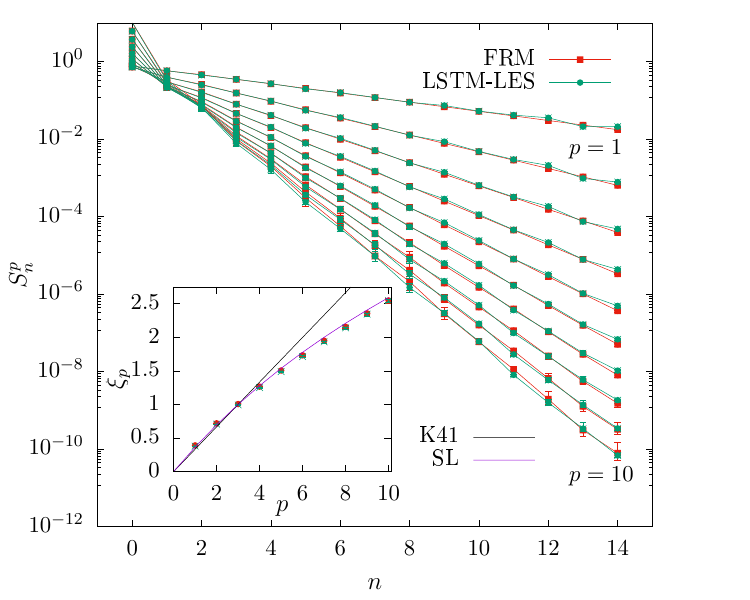}} 
   \caption{Eulerian structure functions $S_n^p = \langle |u_n|^p \rangle$ vs. shell index $n$, in lin-log scale, for orders $p$ from $1$ to $10$ and with $N_{cut}=14$ , comparison between fully resolved model (FRM) and LSTM-LES model.  Inset plot: anomalous scaling exponents $\xi_p$ of Eulerian structure fuctions, $S_n^p  \propto k_n^{-\xi_p}$, for the fully resolved model (FRM), the LSTM-LES model, the prediction from Kolmogorov K41 theory~\cite{K41} and the prediction from She-Leveque
   model (SL)~\cite{SL}. For error-bar estimates see supplementary materials.}
\label{fig:eul}
\end{figure}

\begin{figure}[h]
    \centering
	\resizebox{.5\textwidth}{!}{\includegraphics{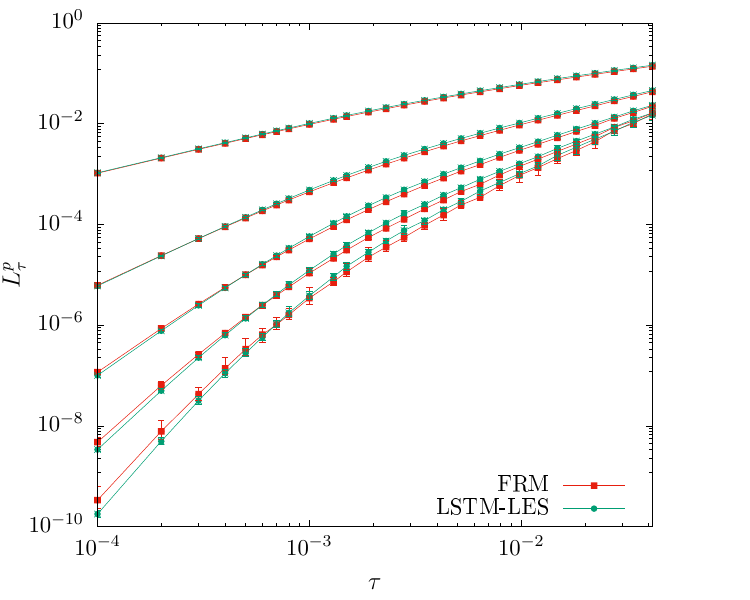}}
    \caption{Lagrangian structure functions (see Eq.~\ref{lagr}), in log-log scale ($\tau$ on the x-axis, $p=1,..,5$), comparison between the fully resolved model (FRM) and LSTM-LES model.  For error-bar estimates see supplementary material.}
    \label{fig:lagr}
\end{figure}

\begin{figure}[h]
        \centering
	\resizebox{.5\textwidth}{!}{\includegraphics{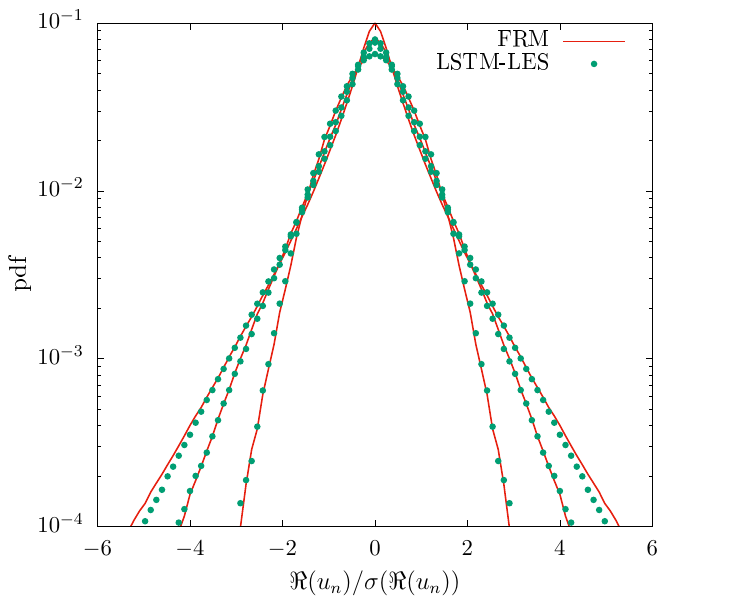}}
	        \caption{PDF of the real part of shells 4, 8 and 12, in log scale,  normalized with standard deviation $\Re u_n / \sigma(\Re u_n)$ for the fully resolved model (FRM), and the LSTM-LES model.} 
        \label{fig:sd}
\end{figure}

\begin{figure}[ht]
		\centering
			\resizebox{.5\textwidth}{!}{\includegraphics{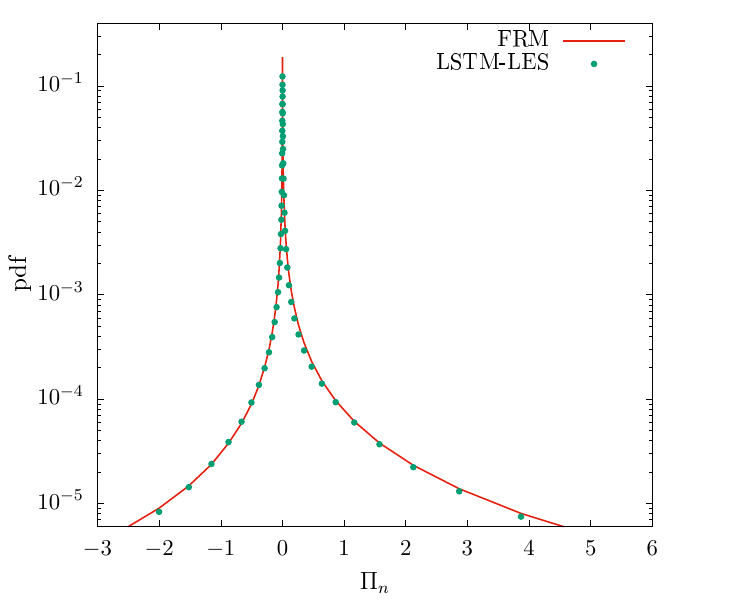}}
        	\caption{PDF of the convective fluxes (Eq.~\ref{flux}) at shell $n=N_{cut}$, for the fully resolved model (FRM) and the LSTM-LES model.}
        	\label{fig:flux}
\end{figure}

This problem has been recently adressed in~\cite{opt}, where it has been formalized in terms of reduced systems of probability distributions, allowing a precise definition of an optimal subgrid model, although practically intractable, and a series of systematic approximations, yet lacking relevant physical features (correct scaling, energy backscatter).  
The approach proposed here, which will be denoted as \textit{LSTM-LES}, employs a custom ODE integrator enhanced by a Long-Short Term Memory (LSTM)~\cite{lstm} Recurrent Neural Network, and is capable of reproducing scaling laws for high order Eulerian (Figure~\ref{fig:eul}) and Lagrangian (Figure~\ref{fig:lagr}) structure functions indistinguishable, within statistical accuracy, from those of the fully resolved model, together with intermittency (Figure~\ref{fig:sd}) and energy backscatter (Figure~\ref{fig:flux}), significantly outperforming the state-of-the-art~\cite{opt}.

\clearpage

We consider the SABRA shell model~\cite{sabra}, for which the governing equations read:
\begin{dmath}
    \frac{d u_n}{dt} = f_n + D_n + C_n = f_n \underbrace{- \nu k_n^2 u_n}_{D_n} +
    i(a k_{n+1} u_{n+2} u^*_{n+1} + b k_{n} u_{n+1} u^*_{n-1} + c k_{n-1} u_{n-1} u_{n-2})
	\label{sabra}
\end{dmath}
with $n=0,1,\dots, N$. For an overview of the parameters, see Table S.I in the Supplementary Material. 
The shell variables $u_n$ evolve following three distinct mechanisms, represented by the three terms in the right-hand side of the equation: the forcing ($f_n$), injecting energy at large scales, the viscous term ($D_n$), dissipating energy at the smaller scales, and the convective term ($C_n$),  a nonlinear coupling between the shells, responsible for the energy cascade mechanism. 
The forcing mechanism chosen for our experiments is constant in time, with the magnitude defined by a parameter $\epsilon$, and it is applied only at large scales, with $f_0 = \epsilon$ and $f_1 =0.7 \epsilon$ (which ensures zero helicity flux~\cite{sabra}), and determines, together with the value of the viscosity $\nu$, the Reynolds number of the model ($Re \sim 1/\nu$). 

We now introduce the two models that will be confronted in this paper namely the Fully Resolved Model (FRM), used to generate large amount of data with converged statistics of high-order statistical moments, and our LSTM-LES model, to be trained on the basis of the generated data, and whose target is to reproduce the statistical behavior of the FRM.

The FRM is resolved with a $4^{th}$ order Runge-Kutta scheme with explicit integration of the viscosity~\cite{dynturb}. The integration timestep $\Delta t_{GT}$ (for \textit{Ground Truth}) needs to be sufficiently small as to resolve the complete energy cascade up to the Kolmogorov scale (shell index $N_{\eta}$).  

The LSTM-LES model, instead, is composed of the same $4^{th}$ order Runge-Kutta scheme for the evolution of the  large scale DOFs $\{ u_n, n=0,\dots,N_{cut}\}$, and a Recurrent Artificial Neural Network, modelling the subgrid closure. The timestep used in this model will be indicated by $\Delta t_{LES}$ and is considerably bigger than $\Delta t_{GT}$ (see Table S.I in the Supplementary Material), being $N_{cut} \ll N_{\eta} $ the smallest scale that needs to be explicitly resolved, significantly reducing the computational costs.

Artificial Neural Networks are a class of complex nonlinear models that have been shown to satisfy, under some hypothesis, universal approximation theorems for arbitrary functions~\cite{uniapprox}. In the last decade they have obtained outstanding results in various fields~\cite{ml2}, also comprising fluid dynamics and turbulence modelling~\cite{mlturb1,mlturb2,repred}. 
We employ a Recurrent Neural Network (RNN) architecture~\cite{rnn}, which is particularly useful in the context of time series data since the input of the neural network is composed by a \textit{present} input term and by a \textit{state} (or memory) term, coming from a previous evaluation of the network. 
In particular, we use Long-Short Term Memory (LSTM) networks, which are a particular istance of recurrent neural networks that implement a gating mechanism in order to alleviate the problem of vanishing/exploding gradients during training~\cite{lstm}. 

Before explaining the functioning of the LSTM-LES model, described extensively in the supplementary materials, we highlight one of the main assumptions made in the definition of the SABRA shell model.  This feature, crucial for our subgrid closure approach, is the locality in Fourier space of the convective term $C_n$, mimicking the Richardson energy cascade model~\cite{frisch}.
The relevance of the locality hypothesis has been extensively discussed for example in~\cite{loc1,loc2}. 
This means that the convective term for the shell with index $n$ depends only on shells indexed by $\{ n-2, n-1, n, n+1, n+2\}$ (see Equation~\ref{sabra}). 
This implies that, after performing the subgrid cut at shell $N_{cut}$, the only shells for which $\frac{d u_n}{dt}$ remains undefined are $n=N_{cut}$ and $n=N_{cut}-1$, and that to compute the value of the derivative for those shells is necessary to model the values of $u_{N_{cut}+1}$ and $u_{N_{cut}+2}$. The subgrid closure modelling, performed by the Neural Network, consists then in the prediction of these two shells given the values of the large scales at the current timestep and the state of the LSTM, from the previous evaluation:
\begin{equation}
\begin{pmatrix} u_0(t), \cdots, u_{N_{cut}}(t) \\ state(t) \end{pmatrix}  \xmapsto{\text{LSTM}} \begin{pmatrix} u_{N_{cut}+1}(t), u_{N_{cut}+2}(t) \\ state(t+\Delta t_{LES}) \end{pmatrix}.
\end{equation}

We train our Neural Network in an a-priori fashion by asking for one-step pointwise prediction, without using the predicted value to advance to the next timestep (for more details on the training procedure, see the supplementary material). We opt for this approach because of the chaoticity of turbulence, reproduced by the SABRA shell model~\cite{shellrev}. Analogously, we test the network by confronting the LSTM-LES model and the FRM only in terms of statistical quantities such as distributions and statistical moments, avoiding meaningless computation of pointwise errors.

In the following, we present and discuss the performances of the LSTM-LES model in comparison with the fully resolved model (FRM) and the literature~\cite{opt}.

We begin by considering the scaling of the $p^{th}$ order Eulerian structure functions, with $p=1,\cdots,10$ computed as:
\begin{equation}
S_n^p = \langle |u_n|^p \rangle.
\label{eul}
\end{equation}
The results are shown in Figure~\ref{fig:eul}. The LSTM-LES model is indistinguishable, within statistical accuracy, from the fully resolved model. This can be further verified by considering the value of the scaling exponents $\xi_p$ ($S_n^p \propto k_n^{-\xi_p}$) as a function of the order, $p$, of the structure function, as shown in the inset of Figure \ref{fig:eul} and reported in Table~\ref{table:anomal}.

\begin{table}[h]
\begin{center}
\begin{tabular}{|c ||c| c| c| c|}
\hline
		Order & K41 & SL & FRM & LSTM-LES \\
                \hline
$1$ & $0.3333$ & $0.3640$ & $0.423 \pm 0.005$ & $0.423 \pm 0.005$ \\ 
$2$ & $0.6667$ & $0.6959$ & $0.78 \pm 0.01$ & $0.78 \pm 0.01$ \\ 
$3$ & $1.0000$ & $1.0000$ & $1.10 \pm 0.02$ & $1.10 \pm 0.02$ \\ 
$4$ & $1.3333$ & $1.2797$ & $1.38 \pm 0.03$ & $1.39 \pm 0.02$ \\ 
$5$ & $1.6667$ & $1.5380$ & $1.64 \pm 0.04$ & $1.65 \pm 0.03$ \\ 
$6$ & $2.0000$ & $1.7778$ & $1.89 \pm 0.06$ & $1.89 \pm 0.05$ \\ 
$7$ & $2.3333$ & $2.0013$ & $2.11 \pm 0.08$ & $2.12 \pm 0.07$ \\ 
$8$ & $2.6667$ & $2.2105$ & $2.33 \pm 0.11$ & $2.33 \pm 0.09$ \\ 
$9$ & $3.0000$ & $2.4074$ & $2.54 \pm 0.14$ & $2.53 \pm 0.12$ \\ 
$10$ & $3.3333$ & $2.5934$ & $2.74 \pm 0.16$ & $2.73 \pm 0.15$ \\
                \hline
\end{tabular}
    \end{center}
	\caption{Anomalous exponents $\xi_p$ for the scaling laws of the Eulerian
		structure fuctions $S_n^p = \langle|u_n|^p\rangle \propto k_n^{-\xi_p}$ for the fully resolved model (FRM), the
		LSTM-LES model, the prediction from Kolmogorov K41 theory,
		and the prediction from She-Leveque model (SL).  For error-bar estimates see supplementary material.}
    \label{table:anomal}
\end{table}

In order to establish the capability of the LSTM-LES model to reproduce time correlations of the signal, we considered the behavior of the Lagrangian structure functions. 
These are computed as the time correlation functions of Lagrangian signals obtained by summing the real parts of all the shells $u(t) = \Re (\sum_n u_n(y))$~\cite{shellrev}: 

\begin{dmath}
L^p_{\tau} = \langle|u(t+\tau)- u(t)|^p \rangle.
\label{lagr}
\end{dmath}
The results are shown in Figure~\ref{fig:lagr}, where we can see that the LSTM-LES model closely follows the scaling laws of the fully resolved model.

In terms of statistical distributions, we show, in Figure~\ref{fig:sd}, the distributions for the real part of shells 4, 8 and 12 ($N_{cut}=14, N_{\eta}=30$), normalized by standard deviation, $(\Re u^n)/(\sigma(\Re u^n))$. Notably, the model correctly reproduces intermittency at the small scales, characterized by the non-gaussian nature of the pdf.

In order to evaluate the capability of the subgrid closure to properly model the energy fluxes to the unresolved scales, we consider the distribution for the convective fluxes computed at the subgrid cut $N_{cut}$~\cite{opt}:

\begin{dmath}
\Pi_{N_{cut}} = k_0 \lambda^{N_{cut}} (\lambda a u_{N_{cut}+2} u_{N_{cut}+1}^* u_{N_{cut}}^* + b u_{N_{cut}+1} u_{N_{cut}}^* u_{N_{cut}-1}^*).
\label{flux}
\end{dmath}
The results are shown in Figure~\ref{fig:flux}, where we can see that the phenomenon of backscatter, i.e.\ the presence of events of net energy fluxes going from small scales to large scales (negative values of the flux in the plot), is well captured by our model. This behavior is tipically very hard to reproduce in typical LES schemes and, when included improperly, it can even compromise the scheme stability~\cite{frisch}. 

\begin{figure}[h]
    \centering
		\resizebox{.5\textwidth}{!}{\includegraphics{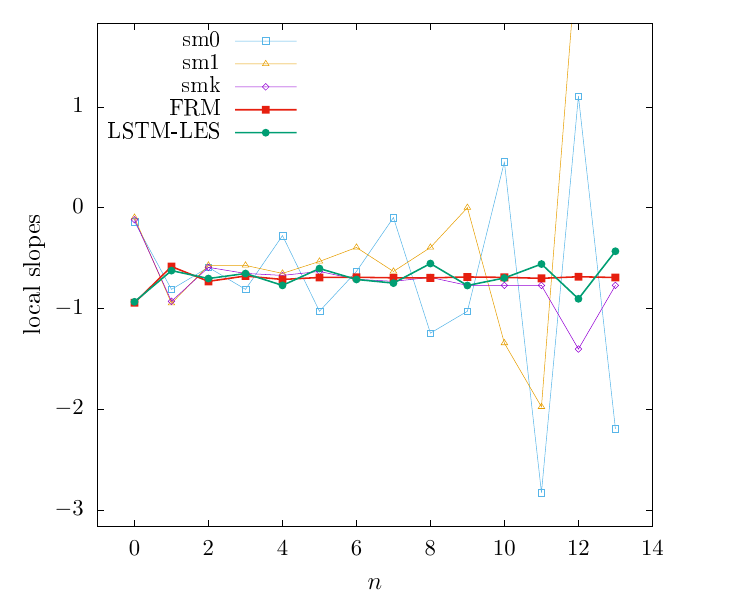}}
		\caption{Local slopes for the second order Eulerian structure function $\frac{\Delta S^2_n}{\Delta n}$ vs. shell index $n$. Comparison between fully resolved model (FRM), LSTM-LES model and three physics based model introduced in~\cite{opt}, denoted as \textit{sm0}, \textit{sm1} and \textit{smk}. } 
        \label{fig:confront}
\end{figure}

Finally, we present a comparison of the LSTM-LES model with the three physics-based alternatives proposed in~\cite{opt}, denoted respectively with \textit{sm0}, \textit{sm1} and \textit{smk}, and corresponding to three different approximation of the optimal subgrid closure scheme.
The comparison is made in terms of local slopes for the $2^{nd}$ order Eulerian structure functions,  computed as finite differences $\frac{\Delta S^2_n}{\Delta n}$ (Figure~\ref{fig:confront}).
We can see that our model significantly outperforms the three models presented in~\cite{opt}. In fact, the models sm0 and sm1 present wide oscillations around the correct scaling law $\frac{\Delta S^2_n}{\Delta n} = -\xi_2 \approx -0.78$. The model denoted as smk instead, i.e.\ the one with better scaling properties, does not entail backscatter, which is present (though not perfectly resolved) in sm0, sm1. 

In this work we have provided a numerical proof of the fact that, in the context of the Shell Models, turbulence closures are possible. Specifically, we have focused on the ambitious goal of developing a model for the small (unresolved) scales capable of reproducing a dynamics of the large (resolved) scales that is statistically indistinguishable from the fully resolved dynamics. Our model, based on a deep learning approach, implements a closure on a shell model of turbulence. While shell models are simplified versions of the Navier-Stokes equation, they are well know for faithfully reproducing the complex phenomenology of the turbulent energy flux, including multiscale correlations and intermittency.

Employing shell models for turbulence allowed us to work in a well-controlled environment where high Reynolds numbers, and thus clean scaling laws over several orders of magnitude, are easily achievable. Additionally, due to the computational lightweight of the shell models, we could perform stringent statistical comparisons, something that would have not been easily achievable in the framework of the full Navier-Stokes equation.

We numerically demonstrated that our subgrid closure model, starting from an initial configuration not present in the training set, generates a dynamics indistinguishable, in statistical terms, from that of a filtered fully resolved simulation. To this aim, we considered high order Eulerian and Lagrangian structure functions together with distributions of shell variables and convective fluxes. Additionally we showed that our model yields results that outperform the state-of-the-art physics based approaches proposed in literature~\cite{opt}.

Our results provide solid foundations towards the possibility to develop subgrid closure for Navier-Stokes that are capable of faithfully reproducing all statistical properties of 3D turbulence.  However, we need to stress that the current work is only a very first step toward the actual development of possible closures for 3d Navier-Stokes. The novelty and value of our results is limited to the finding that the inertial range statistical properties of turbulence (including intermittency in the energy cascade) can be quantitatively reproduced via the sole knowledge of the present state and past state of the resolved scales. Generalizing our approach to  3d Navier-Stokes turbulence still requires overcoming several challenges. These include having to deal with many more subgrid variables to be modeled and by the computational cost of producing enough high-quality 3d training data.

\bibliography{biblio}

\section{Supplementary Materials}
\label{sec:suppl}

\subsection{LSTM-LES model}
We provide here the details about our LSTM-LES deep learning model, implementing the subgrid closure for the SABRA shell model. This is composed by a combination of a $4^{th}$ order Runge Kutta integration scheme with explicit integration of the viscosity~\cite{dynturb}, together with a LSTM Recurrent Neural Network that provides the value of the fluxes to/from the unresolved scales. 

Thanks to the locality of the convective term $C_n$ (Eq. 1), we can
close (and resolve) the large scale system, denoted here by convenience $u^< =
\{u^n, n=0,\cdots,N_{cut}\}$, by only providing the values of the shells below
the subgrid cutoff scale, denoted $u^> = \{u_{N_{cut}+1}, u_{N_{cut}+2}\}$,
given as output of a LSTM Neural Network. However, this is only true for
one-step schemes such as Explicit Euler. As we employ a $4^{th}$ order Runge
Kutta multistepping integration scheme, a dependency on higher shell indexes
(up to $u_{N_{cut}+8}$) emerges, since we need more terms to approximate the
higher order terms in the scheme. Our approach consists then in an  application of the LSTM Neural Network for each of the sub-steps of the multistepping scheme.

\begin{figure}[h]
  \renewcommand\thefigure{S.1}
  \centering
  \includegraphics[width=.5\textwidth]{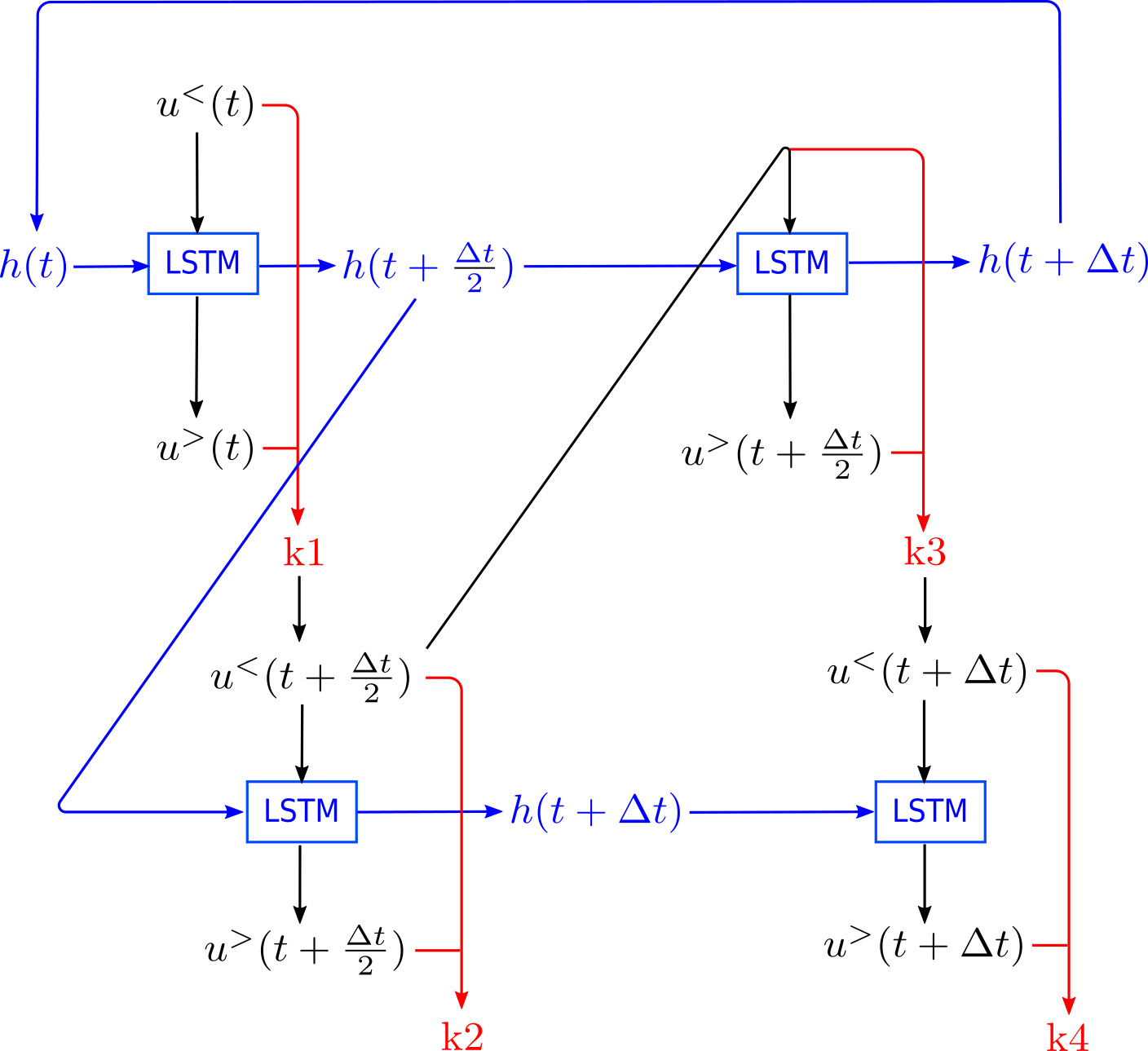}
  \caption{LSTM-LES model. The
    input of the timestepping scheme are the hidden state $h(t)$ and the value of the solution at the large scales,
    before the cut, $u^{<}(t)$. The scheme computes the four terms $\textbf{k}_1,\textbf{k}_2,\textbf{k}_3$
    and $\textbf{k}_4$, and uses those to update $u^{<}(t)$ according to Eq. S.1}
    \label{fig:model}
\end{figure}

Specifically, the LSTM-LES model (cf. Figure~\ref{fig:model}) takes as input the large scales of turbulence, before the subgrid cutoff $N_{cut}$, at time $t$, $u^< (t)$, and the hidden state from the previous timestep $h(t)$, and returns $u^< (t+\Delta t)$  and the hidden state for the next timestep $h(t+\Delta t)$ (here and in the following we will indicate for conciseness $\Delta t_{LES}$ with $\Delta t$).

In order to compute $u^< (t+\Delta t)$, we need to evaluate the four terms $\textbf{k}_1, \textbf{k}_2, \textbf{k}_3$ and $\textbf{k}_4$, and use them to advance the large scales in time, in formulas:

\begin{equation}
  \renewcommand\theequation{S.1}
    u^< (t+\Delta t) = u^<(t) + \frac{\Delta t}{6} (\textbf{k}_1 + 2 \textbf{k}_2 + 2 \textbf{k}_3 + \textbf{k}_4).
    \label{rk}
\end{equation}
The scheme is then composed of 4 steps:
\begin{enumerate}
	\item (top left of Figure~\ref{fig:model}) give $u^<(t)$ and
        $h(t)$ (both from the previous timestep) as input to the LSTM, and take $u^{>}(t)$ 
		and $h(t+\frac{\Delta t}{2})$ as output. Use $u^<(t)$ and $u^>(t)$ 
		to compute $\textbf{k}_1=\frac{d u^<}{dt}(t)$ via
		Equation 1. 
    After that, compute $u^<(t+\frac{\Delta t}{2})$ using an Exlicit Euler scheme:
\begin{equation}
  \renewcommand\theequation{S.2}
    u^< (t+\frac{\Delta t}{2}) = u^<(t) + \frac{\Delta t}{2} \frac{d u^<}{dt}(t);
    \label{ee}
\end{equation}
	\item (bottom left) use $u^< (t+\frac{\Delta t}{2})$ (computed in 1.) 
        and the hidden state
        $h(t+\frac{\Delta t}{2})$ (computed in 1.) to compute 
        $\textbf{k}_2=\frac{d u^<}{dt}(t+\frac{\Delta t}{2})$ and $h(t+\Delta
		t)$;
    \item (top right) use $u^< (t+\frac{\Delta t}{2})$ (computed in 2.) and
        $h(t+\frac{\Delta t}{2})$ (computed in 1.) to compute 
        $\textbf{k}_3=\frac{d u^<}{dt}(t+\frac{\Delta t}{2})$, use that to compute $u^< (t+\Delta t)$, and
        feed in the output hidden state $h(t+\Delta t)$ for the next timestep;
    \item (bottom right) use $u^< (t+\Delta t)$ (computed in 3.) and                
        $h(t+\Delta t)$ (computed in 2.) to compute $\textbf{k}_4=\frac{d u^<}{dt}(t+\Delta t)$.
\end{enumerate}

In the scheme above, the LSTM model used (architecture, values of weights/biases) is the same for all the 4 steps. The only differences are in the input, $u^<$, and the hidden state, $h$, resulting in different outputs. Moreover, the hidden states are organized to have time consistency of the LSTM, i.e.\ the input of the LSTM $u^<$ and the hidden state $h$ are evaluated at the same timestep, where one evaluation of the LSTM corresponds to half a timestep. Finally, we remark that requiring the time consistency between hidden states and solution does not fully define the scheme, and that we could have reorganized the topology of the graph of Figure \ref{fig:model} in different ways, possibly obtaining different results. From the experiments done, different reorganizations of the topology did not give qualitateively different results; this will be investigated more deeply in a future work.

\begin{table}[h]
  \renewcommand\thetable{S.I}
    \begin{center}
		\begin{tabular}{|p{20mm} |p{25mm}| p{35mm}|}
    
    \hline
		\textbf{Parameter} & \textbf{Value} & \textbf{Description}\\
    \hline
		$\nu$ & $1 \times 10^{-12}$ & viscosity \\
    \hline
		$\text{Re}$ & $\approx 10^{12}$ & Reynolds number \\
    \hline
		$\epsilon$  & $0.5$ & forcing \\
    \hline
			$N$ & $40$ & number of shells	\\
    \hline
			$N_{\eta}$ & $30$ & Kolmogorov scale	\\
    \hline
			$N_{cut}$ & $14$ & subgrid cutoff scale \\
    \hline
			$\tau_0$ & $7.553 \times 10^{-1}$ & eddy turnover time for the integral scale \\
    \hline
			$\tau_{\eta}$ & $1.8367 \times 10^{-6}$ & eddy turnover time for the dissipative scale \\
    \hline
			$\Delta t_{GT}$ & $1 \times 10^{-8}$ & timestep of FRM\\
    \hline
			$\Delta t_{LES}$ & $1 \times 10^{-5}$ & timestep of LES-LSTM model\\
    \hline
			$N_{data}$ & $5000$ & number of initial conditions of dataset\\
    \hline
			$N_{batch}$ & $3000$ & batch size for training\\
    \hline
			$T_{train}$ & $20480 \times \Delta t_{LES}$ & integration time of
			training dataset\\
    \hline
			$T_{test}$ & $81920 \times \Delta t_{LES}$ & integration time of
			test dataset\\
    \hline
    			$\text{bptt}$ & $32$ & backpropagation through time truncation parameter\\
    \hline
    \end{tabular}
    \end{center}
	\caption{Values of the parameters of the numerical experiments.}
    \label{table}
\end{table}

\subsection{Model Training and Testing}

The ground truth datasets for training of the LSTM is generated by integrating
the shell model with the FRM model for a certain number of random initial 
conditions $N_{data}$ and for a fixed time interval $T_{train}$ (after a transient). 
The timestep used in this phase is $\Delta t_{GT}$. 
We dump the solution every  $\frac{\Delta t_{GT}}{\Delta t_{LES}}$ timesteps, where $\Delta t_{LES}$ is
the timestep of the LSTM-LES model, for a total of $T_{train}$ time
snapshots.

The training is then done by considering a batch size of $N_{batch}$ intial conditions, randomly sampled at each epoch from the training set, and by performing Truncated Backpropagation Through Time (TBTT) over it~\cite{rnn}. 

Specifically, we take the following steps:

\begin{algorithmic}
\State $epoch \gets 0$
\While{$epoch < n\_epochs$}
	\State $t \gets 0$
	\State select randomly $N_{batch}$ ICs from $N_{data}$ $\rightarrow$ $u^<(0)$
	\While{$t<T$}
		\State $loss \gets 0$
		\For{$k=0$; $k<\text{tbtt}$; $k \mathrel{+}= 1$ }
			\State $u^<(t+\Delta t_{LES}) = \text{LSTM-LES}(u^<(t)) $
			\State $t \gets t + \Delta t_{LES}$ 	
			\State $loss \gets loss + \mathcal{L}(u^<(t),u_{GT}^<(t))$
		\EndFor
		\State backpropagate loss through LSTM
		\State update LSTM parameters via gradient descent
	\EndWhile
\EndWhile
\end{algorithmic}

As a loss function, $\mathcal{L}(u,v)$, we used the classical $L^2$ loss $||u-v||^2$ applied to shells $N_{cut}-8$ to $N_{cut}$. For the gradient descent we used the Adam optimizer with adaptative learning rate (initial value $lr_0=3e-4$).

The testing is done by confronting the FRM model and the LSTM-LES model in terms of statistical quantities (Eulerian and Lagrangian structure functions, distributions, anomalous exponents). These statistics are computed for the two models, respectively:
\begin{itemize}
\item (FRM model) by considering the test ground truth dataset, i.e.\  $N_{data}$ different initial conditions for $T_{test}$ timesteps. The timestep used in this phase is $\Delta t_{GT}$. 
We dump the solution every  $\frac{\Delta t_{GT}}{\Delta t_{LES}}$ timesteps, where $\Delta t_{LES}$ is
the timestep of the LSTM-LES model.

\item (LSTM-LES model) by taking the final state of the $N_{data}$ solutions of the test ground truth dataset and integrating forward, using the LSTM-LES model, for the same number of timesteps $T_{test}$ of size $\Delta t_{LES}$.
\end{itemize}

We estimate the error-bars by splitting the datasets in chunks consisting of $1024$ time snapshots (for a total of $80$ chunks). We then compute individual statistics on these chunks, and report the average of these statistics as a central point, and the difference between minimum and maximum as the error-bar.

\end{document}